\documentclass{ab}
\usepackage{graphicx}
\usepackage{natbib}

\begin{document}  


\title{Data Acquisition and Control System for High-Performance
Large-Area CCD Systems}

\author{{I.~V.}~{Afanasieva}}

\institute{Special Astrophysical Observatory, Russian Academy of Sciences, Nizhnij Arkhyz, 369167, Russia}

\titlerunning{Data acquisition and control system}
\authorrunning{Afanasieva}

\date{February 16, 2015/Revised March 27, 2015}
\offprints{Irina Afanasieva \email{riv@sao.ru}} 

\abstract{
Astronomical CCD systems based on second-generation DINACON
controllers were developed at the SAO~RAS Advanced Design
Laboratory more than seven years ago and since then have been in
constant operation at the 6-meter and Zeiss-1000 telescopes. Such
systems use monolithic large-area CCDs. We describe the software
developed for the control of a family of large-area CCD systems
equipped with a DINACON-II controller. The software suite serves
for acquisition, primary reduction, visualization, and storage of
video data, and also for the control, setup, and diagnostics of
the CCD system.
}

\maketitle

\section{INTRODUCTION}

The Advanced Design Laboratory of the Special Astrophysical
Observatory of the Russian Academy of Sciences develops and makes
CCD systems (CCD-based digital image acquisition systems) for the
Eurasia's largest 6-m  telescope and other optical telescopes. The
system~\citep{mar00:Afanasieva_n} consists of a CCD array detector,
incorporated into a cryostatic camera, and a CCD controller with
an interface for digital data input into a remote control computer
(the host computer). We now focus on image acquisition systems
equipped with the E2V~CCD\,42-40 (\mbox{2048$\times$2048}~pixels)
and E2V~\mbox{CCD\,42-90} (\mbox{2048$\times$4608}~pixels)
detectors. The data are fed to the control computer via a 100~Mbit
standard Ethernet unit. The software required for the control of
the CCD system and acquisition of astronomical data (images)
should be reliable, ensure high throughput, and provide tools for
setup and diagnostics of the data acquisition system.
Object-oriented methodologies and developing and debugging
tools~\citep{goma02:Afanasieva_n} allow rapid development of such
software suites.

\section{ARCHITECTURE OF THE DATA ACQUISITION SYSTEM}

The development of the program was carried out in line with the requirements it was supposed to meet.
The acquisition system is to be able to perform the following functions:
\begin{itemize}

\item provide interface between the user and the CCD system;

\item run the controller;

\item ensure data acquisition and storage;

\item visualize and analyze the data;

\item perform setup, telemetry, and diagnostics of the CCD system;

\item test the photoelectric characteristics of the CCD system.
\end{itemize}
The system should also allow programmable setting of the integration and readout parameters
including:
\begin{itemize}

\item exposure time;

\item shutter control parameters;

\item parameters of detector flushing before charge integration;

\item image readout speed;

\item number of the detector output node;

\item size and coordinates of the region of interest (ROI) to
read;

\item binning factor along the two coordinates;

\item video channel gain;

\item exposure mode (frame-by-frame, combined, scanning).
\end{itemize}
The program architecture should combine end-user requirements
with technical specifications via understanding of the use cases with subsequent finding of
the ways of their implementation~\citep{mac02:Afanasieva_n}.

We formulated the main quality criteria to be met by the architecture of the acquisition and control
system:
\begin{itemize}

\item {\it universality}, the ability to adapt to controlling CCDs
of various types used in astronomy and to interact with various
digital data input interfaces;

\item {\it flexibility}, implementation of all integration and
readout modes that are possible for the detector of the type
considered, full programmability of CCD control parameters:
voltage levels, clock sweeps, time base, and detector temperature,
and their telemetric control;

\item {\it efficiency}, minimum image readout, recording,
analysis, and visualization times, synchronization of data
exchange streams, ensuring the efficient work of the operator in
the process of data acquisition, and operation of the facility in
the mode of interacting concurrent tasks;

\item {\it adaptability}, implementation of different observing
methods with the possibility of connecting and controlling of
external devices (shutter, filters) from within the acquisition
system;

\item {\it usability}, development of a unified and intuitive user
interface with an extended structure of menus, dialog boxes, and
graphical windows; distributing functions among subsystems,
minimization of data flows between them, possibility of automating
the observing process.
\end{itemize}

The proposed software architecture  (Fig.~1) is based on the
client/server model, which meets the operation efficiency and
reliability
requirements~\citep{mena00:Afanasieva_n,les04:Afanasieva_n,bon04:Afanasieva_n}
and has prospects for future development. The server directly
interacts with the detecting system, however, the actual control
is performed by the clients, which can  connect to the server.

\begin{figure}[tbp!!!]
\includegraphics[width=6.6cm]{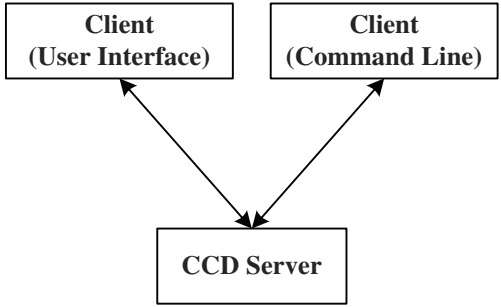}
\caption{Client--server architecture of the control system.}
\label{fig1:Afanasieva_n}
\end{figure}

\begin{figure}[tbp!!!]
 \vspace{1mm}
\includegraphics[width=66mm]{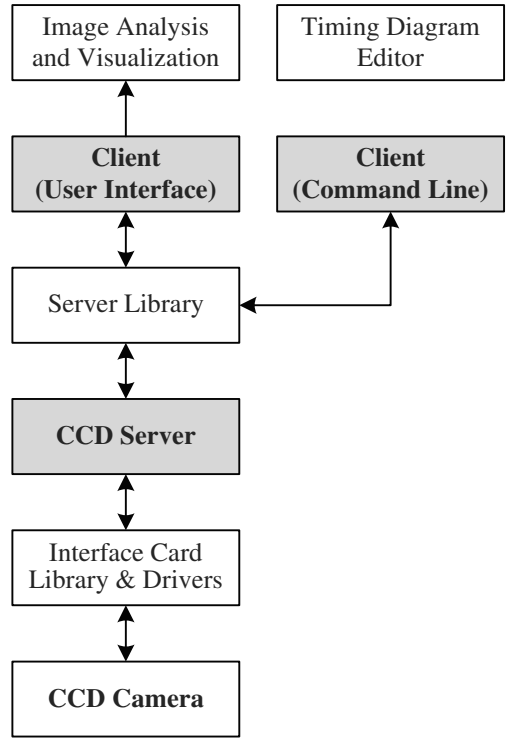}
\caption{Modular architecture of the control system.}
\label{fig2:Afanasieva_n}
\end{figure}

The server and client parts of the system can be hosted on
different physical nodes using network programming technologies or
special client-server interaction middleware. For convenience, let
us refer to clients of the first and second type as graphical and
symbolic respectively. The functions to be performed by the server
part include:
\begin{itemize}

\item  initialization of the CCD system, implementation of various
exposure modes;

\item  setup, testing, and diagnostics of the CCD  system;

\item  interaction with the interface board (Ethernet adapter),
synchronization of reception and transmission of IEEE\,802.3
network packets;

\item  interaction with clients, synchronization of the
acquisition of commands and transmission of the results of their
execution;

\item  formation and recording of image files.
\end{itemize}

Functions of the graphical client:
\begin{itemize}
\item  graphical user interface;

\item  interactive  control of the CCD system;

\item  interface for connecting to and controlling external
devices (shutter, filters);

\item  setting various observing methods;

\item  setting and telemetric control of the detector voltage
levels and temperature;

\item  interactive control of the interface board.
\end{itemize}
Functions of the symbolic client:
\begin{itemize}
 \item control
of the CCD  system from a command line or a batch file.
\end{itemize}

\begin{figure*}[tbp!!!]
\includegraphics[width=170mm]{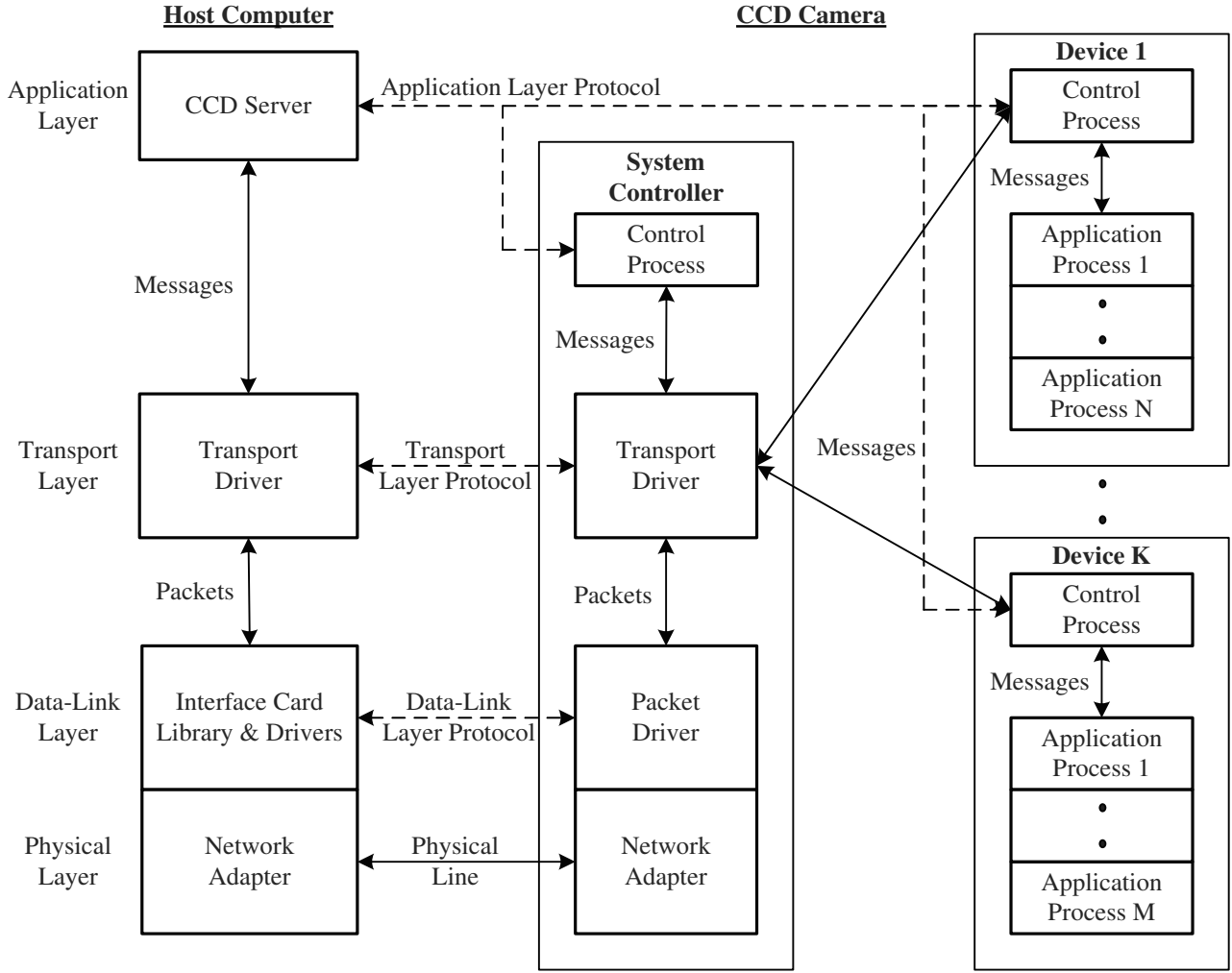}
\caption{Network protocol architecture.} \label{fig3:Afanasieva_n}
\end{figure*}

Visualization of images and editing of time base diagrams can be
implemented in separate applications and used independently of the
server and clients.

Let us now define more exactly the resulting architecture by subdividing it into units
(Fig.~2). Here by a unit we mean a set of closely interconnected
classes and/or objects, subsystems, dependences, operations,
events, and constraints together with a sufficiently well-defined and, as far as possible, compact interface
with other units.

We use the UML language~\citep{fau99:Afanasieva_n} to create
classes and objects and describe the processes and architecture of
the software suite. See our earlier paper for the justification of
this choice~\citep{afa10:Afanasieva_n}.

Let us now consider in more detail the data exchange between the
server and the CCD controller. Because of the need for data
exchange between the control process in the host computer and
application processes in the controller, the underlying principles
resemble those of the corresponding network exchange processes in
many respects. Four layers of the Open Systems Interconnection
(OSI) model~\citep{gos99:Afanasieva_n} are used to create the
interface with the CCD system. The choice of the required layers
is determined by the configuration of the data links in the CCD
system, performance and reliability requirements. The upper
(application) layer ensures independent message exchange between
various processes in the controller on the one part and the host
computer on the other part. The interaction between the  CCD
server and the controller should occur on two (application and
transport) layers. The physical layer and a part of the data link
layer are hardware implemented in the network adapter. Figure~3
shows the network interface hierarchy. Data exchange occurs
through a fiber-optical link via an  Ethernet adapter using the
WinPCAP low-level library for interaction with network device
drivers.\footnote{\tt http://www.winpcap.org/docs/docs\underline{~}412/}

At the application layer, the processes exchange messages.
Messages sent by the host computer to the controller contain
asynchronous control commands, which may include the necessary
data. In the opposite direction data-containing messages (which
include information about the state of the controller) and service
requests are sent. In addition to asynchronous control commands,
the controller also executes synchronous commands that are passed
to it in advance in the form of a data array.

Asynchronous control commands can be subdivided into the following groups:
\begin{itemize}

\item status poll command;

\item power on/off and reset commands;

\item data array exchange commands;

\item commands for the start/stop of controller processes;

\item host computer and controller synchronization commands;

\item command of asynchronous control of external devices.
\end{itemize}
Synchronous commands include the following groups:
\begin{itemize}

 \item commands
of the synchronous control of charge integration/transfer;

\item synchronous readout control commands;

\item commands of synchronous control of external devices;

\item instruction execution sequence control command.
\end{itemize}
At the transport layer the computer and CCD controller exchange
Ethernet IEEE 802.3 packets\linebreak (frames). The following
commands provide the interface between the transport and
application layers:
\begin{itemize}
 \item connect
({\tt TransportConnect});

\item disconnect ({\tt TransportDisconnect});

\item write message ({\tt TransportWriteMsg});

\item read message ({\tt TransportReadMsg});

\item read transport status ({\tt TransportStatus});

\item reset transport level ({\tt TransportReset}).
\end{itemize}
Given that the clients and the server are independently running
applications, a proper interprocess communication (IPC) method
should be chosen. We use a very simple an easy-to-implement data
streaming method---a named pipe, i.e., a dedicated data link
connecting the two processes. One of the processes creates the
channel and the other one opens it, after which both processes can
send the data via this channel in one or both directions using
file read/write functions. This method makes it possible to
organize data exchange both between the local processes and
between  processes started on different network nodes. Client
commands can be of two types: serving for (1) the interaction with
the server exclusively (information commands) and (2) for the
interaction with the controller. The data of the commands of the
first type are sent via channels {\tt C\_PIPE\_INFO} (read) and
{\tt S\_PIPE\_INFO} (write), and those of the commands of the
second type---via channels {\tt C\_PIPE} (read) and {\tt S\_PIPE}
(write). All channels are created by the server. Applications use
semaphores, critical sections, mutexes, queues, and messages for
synchronizing tasks and intertask data exchange.

The main characteristic of the CCD server is its control state, which identifies
the working program segment and corresponds to the current state of the CCD system.
The server is in the initial state (Standby) when it starts. The system
considered has a total of six control states (Fig.~4). Transitions between the states
are triggered by client commands ({\tt
setup, observe, stop, abort, and run~cmd}).

\begin{figure}[tbp!!!]
\includegraphics[width=76mm]{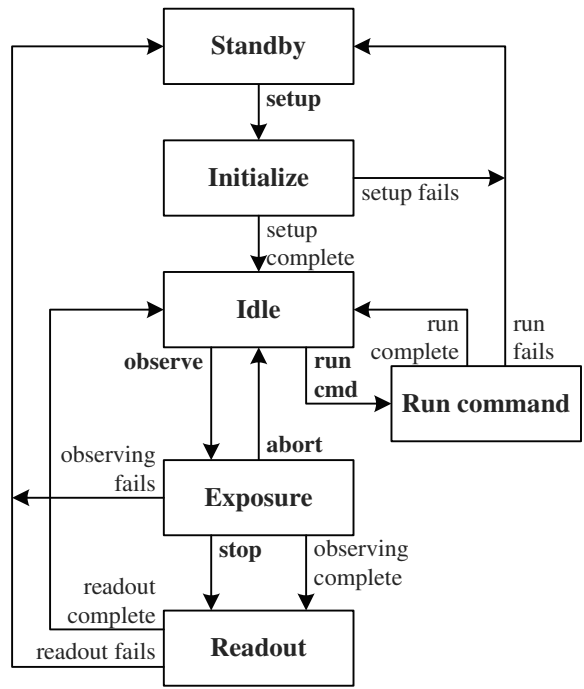}
\caption{Control states of the CCD server.}
\label{fig4:Afanasieva_n}
\end{figure}

\begin{figure*}[tbp!!!]
\centerline{\includegraphics[width=160mm]{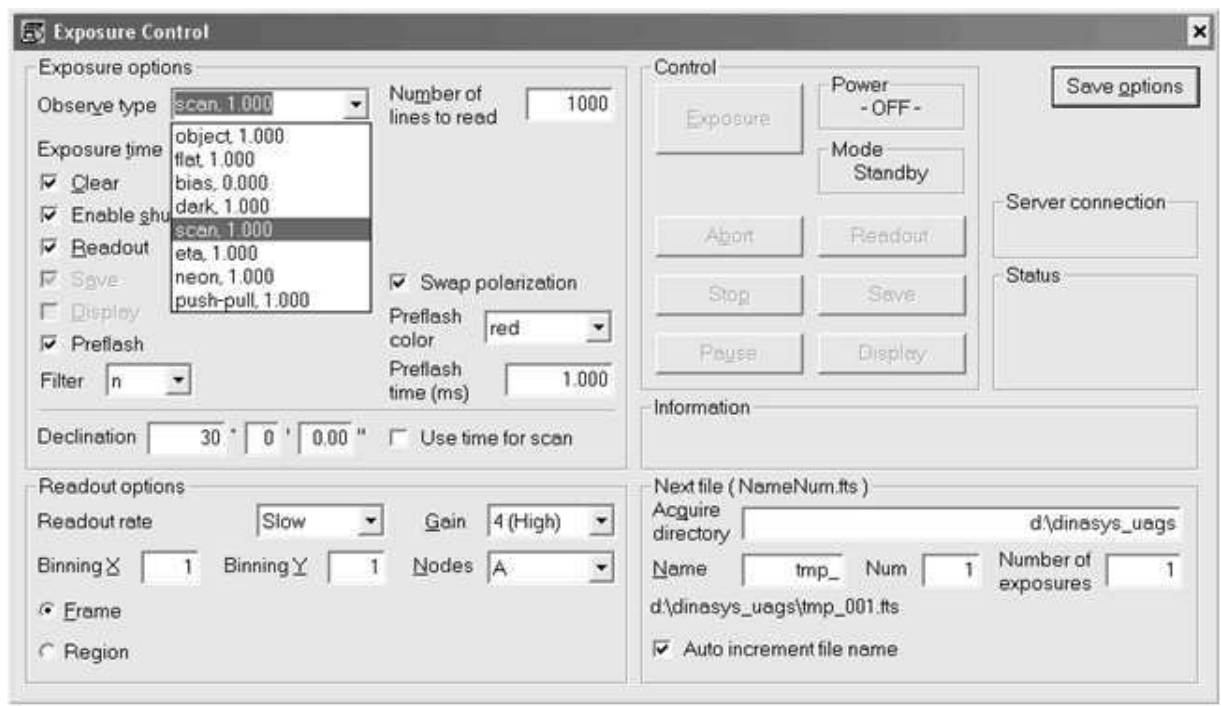}}
\caption{Exposure Control dialog box.} \label{fig5:Afanasieva_n}
\end{figure*}

\section{USER INTERFACE AND CAPABILITIES OF~THE~SYSTEM}

The software suite that we developed runs in the Windows~XP
operating system, has a convenient and intuitive user interface
and an extended menu system. Its multimodule and multitask
architecture allows the astronomer to  perform several tasks
simultaneously both in the interactive and automatic modes. The
graphical client provides the following control and telemetry
windows:
\begin{itemize}

\item Exposure
Control, to control the CCD system;

\item Temperature, for the setup and telemetry of the detector
temperature;

\item Wave Levels, for the setup and telemetry of detector clock
levels;

\item Output Node Levels, for the setup and telemetry of the
working voltages and currents of the output stages of the
detector.
\end{itemize}
The  Exposure Control dialog box (Fig.~5) allows the user to
efficiently control the detector, select various image integration
and readout methods, and set the exposure parameters. It is used
to set the following parameters: exposure type, integration time,
flag of detector flushing before integration, shutter flag, number
of the filter, readout speed, binning ratio along the two
coordinates, video channel gain, output nodes of the detector,
size and coordinates of the ROI to read, number of exposures,
parameters of visualization and writing to a file. This dialog box
also displays information about the progress of command execution,
the controller power and control states. Acquisition of images
with different readout speeds reduces substantially the time spent
to prepare the system for observations. Data acquisition can be
started after successful initialization of the CCD system.
Exposure starts with a click of the Exposure button. After
integration the data are digitized, transferred to the computer,
and written to a file on the disk.

The observer can choose from the following exposure  types:
\begin{itemize}

\item  {\tt bias}, the signal is read from the CCD and then
digitized without counting time and opening the shutter;

\item  {\tt dark}, the charge is accumulated during predetermined
time interval, the signal is read from the CCD and digitized, the
shutter is not opened;

\item  {\tt object}, integration and readout in the frame-by-frame
mode; the shutter is opened, charge is accumulated during certain
time interval, the shutter is closed, the signal is read from the
CCD and digitized;

\item  {\tt neon}, the selected filter is installed, the
comparison spectrum lamp is switched on, the shutter is opened,
charge is accumulated over a certain time interval, the comparison
spectrum lamp is switched off, the signal is read from the CCD and
digitized;

\item  {\tt scan}, integration and readout proceed in the
drift-scan mode; the shutter is opened, the first data row of the
CCD is read and digitized, after certain time the second row is
read and so the given number of rows are read one by one, and the
shutter is closed; the drift-scan mode allows long observations of
sky objects to be performed;

\item {\tt push-pull}, the combined integration and transfer
mode~\citep{cho00:Afanasieva_n}; in addition to standard exposure
parameters, the user sets the elementary exposure time, number of
transfers, and the number of rows by which the image is
transferred.
\end{itemize}

The built-in plug-in mechanism allows any additional devices (filter turrets,
comparison spectrum lamps, etc.) to be connected to the CCD system.
A full-blown internal batch-programming macro language allows automating individual
processes of the control of image integration, readout, and visualization.
The system also provides:
\begin{itemize}
\item a  data acquisition function for computing the
transformation quantum and readout noise of video channels;

\item a  data acquisition function for measurement the noise
spectra of the video channel in order to determine the
coefficients of the optimum filter used during real-time digital
video processing in the video channel;

\item a function for the acquisition of the transfer
characteristic of video channels for its subsequent digital
correction in order to stabilize and linearize it.
\end{itemize}

\section{CONCLUSIONS}
A data acquisition system for optical observations on the 6-meter
and Zeiss-1000 telescopes  has been developed, implemented, and is
currently supported at the Special Astrophysical Observatory.
While in operation, the system proved its relevance and
efficiency. It ensures data acquisition, recording, primary
reduction, and visualization. We have implemented the important
functionality of automating the observing process  and the control
of CCD system parameters during observations.

\begin{acknowledgements}
This work was supported in part by the\linebreak Russian
Foundation for Basic Research (grant\linebreak No.~08-07-12099)
and the Division of Physical Sciences of the Russian Academy of
Sciences (the program {\it Extended Objects in the Universe}).
I~am grateful to all my colleagues who took part in discussing
this work. Observations on the \mbox{6-m} telescope are supported
by the Ministry of Education\linebreak and Science of the Russian
Federation\linebreak (contract No.~14.619.21.0004, project
identifier\linebreak RFMEFI61914X0004).
\end{acknowledgements}

\end{document}